# One Dimensional p-adic Integral Value Transformations

Sk. S. Hassan[1], A. Roy[1], P. Pal Choudhury[1], B.K. Nayak[2]

[1]Applied Statistics Unit, Indian Statistical Institute, Kolkata 700108, India

[2]P. G. Dept. of Mathematics, Utkal University, Bhubaneswar 751004, India

Emails: *sarimif@isical.ac.in*, *ananyaaroy1@gmail.com*, *pabitra@isical.ac.in*,

*bknatuu@yahoo.co.uk*

Correspondence should be addressed to *sarimif@isical.ac.in*

### Abstract

*In this paper, a set of transformations $T^{p,k}$ is defined on $\mathbb{N}_0^K$ to $\mathbb{N}_0$. Some basic and naïve mathematical structure of $T^{p,1}$ is introduced. The concept of discrete dynamical systems through IVT and some further research scope of IVTs are highlighted.*

1. **Introduction:** There are $p^{p^k}$ number of p-states, k-neighbourhood *Cellular Automata* (CA) rules [1, 2, 3, and 4]. A class of discrete transformations on $\mathbb{N}_0^K$ named as *Integral Value Transformations* (IVT), corresponding to each of those CA rules, is introduced in the formation of k-dimensional fractal sequences. Some of the algebraic properties of the set of IVTs are formulated. The notion of differentiability of these discrete transformations is derived. The concept of discrete dynamical systems where IVTs are treated as dynamical system map is introduced.

2. ***Notion of Integral Value Transformation (IVT)***:

The Integral Value Transformations (IVTs) from $\mathbb{N}_0^K$ to $\mathbb{N}_0$ defined as

$IVT^{p,k}{}_j : \mathbb{N}_0^K \to \mathbb{N}_0$
$IVT^{p,k}{}_j((n_1, n_2, \dots n_k)) =$
$\left(f_j(a_0^{n_1}, a_0^{n_2}, \dots, a_0^{n_k})\, f_j(a_1^{n_1}, a_1^{n_2}, \dots, a_1^{n_k}) \dots \dots f_j(a_{l-1}^{n_1}, a_{l-1}^{n_2}, \dots, a_{l-1}^{n_k})\right)_p = m$
where $n_1 = (a_0^{n_1} a_1^{n_1} \dots a_{l-1}^{n_1})_p$, $n_2 = (a_0^{n_2} a_1^{n_2} \dots a_{l-1}^{n_2})_p, \dots\dots n_k = (a_0^{n_k} a_1^{n_k} \dots a_{l-1}^{n_k})_p$
$f_j : \{0, 1, 2, \dots, p-1\}^k \to \{0, 1, 2, \dots, p-1\}$.
m is the decimal conversion from the p adic number.

Let us fix the domain of IVTs as $\mathbb{N}_0$ (k=1) and thus the above definition boils down to the following:

$$IVT^{p,1}{}_j(x) = \left(f_j(x_n)\, f_j(x_{n-1}) \dots \dots f_j(x_1)\right)_p = m$$

where m is the decimal conversion from the p adic number, and $x = (x_n\, x_{n-1} \dots\dots x_1)_p$.

Now, let us denote the set of $IVT^{p,1}{}_j$ as

$$T^{p,1} = \left\{ IVT^{p,1}{}_j : \mathbb{N}_0 \to \mathbb{N}_0 \;\middle|\; \begin{array}{c} 0 \leq j < p^p,\; IVT^{p,1}{}_j(x) = \left(f_j(x_n)\, f_j(x_{n-1}) \dots \dots f_j(x_1)\right)_p = m \\ \text{where m is the decimal conversion from the p adic number} \\ \text{and } x = (x_n\, x_{n-1} \dots \dots x_1)_p \end{array} \right\}$$

In the next section, we are about to explore some basic algebraic structures on $T^{p,1}$.

### 2.1. An alternative definition of IVT:

Let us define a function $F: B_p{}^n \to B_p{}^n$ as $F(x_1, x_2, x_3, \ldots, x_n) = (y_1, y_2, y_3, \ldots, y_n)$

Here we define the set $B_p$ as a set $\{0, 1, 2, \ldots, p-1\}$ and $x_i, y_i \in B_p$ such that $y_i = f(x_i)$ where $f$ is one of $f_j$ which was defined earlier in subsection 2.1.

It is a fact that, $\{1\} \times B_p{}^{n-1} \subseteq B_p{}^n$

So, the above function can be defined as $F: \{1\} \times B_p{}^{n-1} \subseteq B_p{}^n \to B_p{}^n$

If x is an integer with p-adic representation $(1, x_2, x_3, \ldots, x_n)_p$ and $F(1, x_2, x_3, \ldots, x_n) = (y_1, y_2, y_3, \ldots, y_n)$.

In this case F is same as IVT in one dimension of p-adic system.

***Corollary- 1***: Conversion of an integer (non-negative) to a p-adic number (minimal significant digit in p-adic base) is a function given as $C: \mathbb{N}_0 \to \cup_{n \in \mathbb{N}_0} B_p{}^n$ and consequently $C^{-1}: \cup_{n \in \mathbb{N}_0} B_p{}^n \to \mathbb{N}_0$ is also a function.

Therefore, $C^{-1}FC$, a map from $\mathbb{N}_0$ to itself is equivalent to $IVT^{p,1}{}_j$.

Now let us see some of the algebraic structure on $T^{p,1}$ in the subsequent section.

## 3. Some Algebraic Results on $T^{p,1}$

**Theorem: 1** $(T^{p,1}, \oplus, \otimes)$ forms a Commutative Ring under the operations $\oplus$ and $\otimes$ defined as

$(IVT^{p,1}{}_{j_1} \oplus IVT^{p,1}{}_{j_2})(x) = \left(f_{j_1}(x_n) \oplus_p f_{j_2}(x_n)\ f_{j_1}(x_{n-1}) \oplus_p f_{j_2}(x_{n-1}) \ldots \ldots f_{j_1}(x_1) \oplus_p f_{j_2}(x_1)\right)_p$ and

$(IVT^{p,1}{}_{j_1} \otimes IVT^{p,1}{}_{j_2})(x) = \left(f_{j_1}(x_n) \otimes_p f_{j_2}(x_n)\ f_{j_1}(x_{n-1}) \otimes_p f_{j_2}(x_{n-1}) \ldots \ldots f_{j_1}(x_1) \otimes_p f_{j_2}(x_1)\right)_p$

where $x = (x_n\ x_{n-1} \ldots \ldots x_1)_p$ and $\oplus_p$ denotes addition modulo p and $\otimes_p$ denotes multiplication modulo p.

**Proof:** Clearly, $T^{p,1}$ is closed under the operation $\oplus$ since $(IVT^{p,1}{}_{j_1} \oplus IVT^{p,1}{}_{j_2})(x) =$
$\left(f_{j_1}(x_n) \oplus_p f_{j_2}(x_n)\ f_{j_1}(x_{n-1}) \oplus_p f_{j_2}(x_{n-1}) \ldots \ldots f_{j_1}(x_1) \oplus_p f_{j_2}(x_1)\right)_p$

$= \left(f_{j_3}(x_n)\ f_{j_3}(x_{n-1}) \ldots \ldots f_{j_3}(x_1)\right)_p = IVT^{p,1}{}_{j_3}$ for some $0 \leq j_3 \leq p^p$.

Associativity of $\oplus$ follows from associativity of $\oplus_p$.

$IVT^{p,1}{}_0(x) = \left(f_0(x_n)\ f_0(x_{n-1}) \ldots \ldots f_0(x_1)\right)_p$ is the additive identity as $IVT^{p,1}{}_0 \oplus IVT^{p,1}{}_j = IVT^{p,1}{}_j \oplus IVT^{p,1}{}_0 = IVT^{p,1}{}_j$.

For every $IVT^{p,1}{}_j \in T^{p,1}{}_\#$, there exists $IVT^{p,1}{}_k \in T^{p,1}{}_\#$ such that $IVT^{p,1}{}_j \oplus IVT^{p,1}{}_k = IVT^{p,1}{}_k \oplus IVT^{p,1}{}_j = IVT^{p,1}{}_0$ which too follows from the fact that the inverse of every element exists under the operation addition modulo p.

Further, $\oplus$ follows commutativity from commutativity of $\oplus_p$.

Thus, $(T^{p,1}, \oplus)$ forms an abelian group.

$(T^{p,1}, \otimes)$ forms a semi group. Indeed, the operation $\otimes$ is closed since

$(IVT^{p,1}_{j_1} \otimes IVT^{p,1}_{j_2})(x) = \left(f_{j_1}(x_n) \otimes_p f_{j_2}(x_n) \; f_{j_1}(x_{n-1}) \otimes_p f_{j_2}(x_{n-1}) \ldots \ldots \ldots f_{j_1}(x_1) \otimes_p f_{j_2}(x_1)\right)_p$

$= \left(f_{j_4}(x_n) \; f_{j_4}(x_{n-1}) \ldots \ldots \ldots f_{j_4}(x_1)\right)_p$ for some $0 \leq j_3 \leq p^p$.

Associativity of $\otimes$ holds due to associativity of $\otimes_p$. Further, the distributive laws hold good.

Hence the result holds.

Now let us see if $T^{p,1}_{\#}$ forms a vector space under any suitable operations which would give us a better idea of the space of IVT's.

**Theorem: 2** $(T^{p,1}_{\#}, \oplus, \wedge)$ *forms a Vector space over a field* $\mathbb{F}_p$ (*p is prime*) *where* $\wedge$ *denotes scalar multiplication defined as* $(c \wedge IVT^{p,1}_j)(x) = \left(c \otimes_p f_j(x_n) \; c \otimes_p f_j(x_{n-1}) \ldots \ldots \ldots c \otimes_p f_j(x_1)\right)_p$

*where* $x = (x_n \; x_{n-1} \ldots \ldots x_1)_p$ *and* $\otimes_p$ *denotes multiplication modulo p.*

*Proof:*

We know that $(T^{p,1}, \oplus)$ is an abelian group as shown in Theorem-1.

$(c \wedge IVT^{p,1}_j)(x) = \left(c \otimes_p f_j(x_n) \; c \otimes_p f_j(x_{n-1}) \ldots \ldots \ldots c \otimes_p f_j(x_1)\right)_p = \left(f_{j_5}(x_n) \; f_{j_5}(x_{n-1}) \ldots \ldots \ldots f_{j_5}(x_1)\right)_p$ for some $0 \leq j_5 \leq p^p$

Thus $c \wedge IVT^{p,1}_j \in T^{p,1}$ for every $c \in \mathbb{F}_p$ and $IVT^{p,1}_j \in T^{p,1}$. Thus $T^{p,1}$ is closed under scalar multiplication.

$((c_1 \oplus_p c_2) \wedge IVT^{p,1}_j)(x) = \left((c_1 \oplus_p c_2) \otimes_p f_j(x_n) \; (c_1 \oplus_p c_2) \otimes_p f_j(x_{n-1}) \ldots \ldots \ldots (c_1 \oplus_p c_2) \otimes_p f_j(x_1)\right)_p$

$= ((c_1 \otimes_p f_j(x_n)) \oplus_p (c_2 \otimes_p f_j(x_n))) \ldots \ldots \ldots ((c_1 \otimes_p f_j(x_1)) \oplus_p (c_2 \otimes_p f_j(x_1)))_p$

$= \left(c_1 \otimes_p f_j(x_n) \; \ldots \ldots \ldots c_1 \otimes_p f_j(x_1)\right)_p \oplus \left(c_2 \otimes_p f_j(x_n) \; \ldots \ldots \ldots c_2 \otimes_p f_j(x_1)\right)_p$

$= (c_1 \wedge IVT^{p,1}_j)(x) \oplus (c_2 \wedge IVT^{p,1}_j)(x)$ for every $c_1, c_2 \in \mathbb{F}_p$ and $IVT^{p,1}_j \in T^{p,1}$

For $c \in \mathbb{F}_p$ and $IVT^{p,1}_{j_1}, IVT^{p,1}_{j_2} \in T^{p,1}$

$(c \wedge (IVT^{p,1}_{j_1} \oplus IVT^{p,1}_{j_2}))(x) = (c \wedge IVT^{p,1}_{j_1})(x) \oplus (c \wedge IVT^{p,1}_{j_2})(x)$.

$1 \wedge IVT^{p,1}_j = IVT^{p,1}_j$.

$((c_1 \otimes_p c_2) \wedge IVT^{p,1}_j)(x) =$
$\left(((c_1 \otimes_p c_2) \otimes_p f_j(x_n)) \; ((c_1 \otimes_p c_2) \otimes_p f_j(x_{n-1})) \ldots \ldots \ldots ((c_1 \otimes_p c_2) \otimes_p f_j(x_1))\right)_p$

$= \left((c_1 \otimes_p (c_2 \otimes_p f_j(x_n))) \; (c_1 \otimes_p (c_2 \otimes_p f_j(x_{n-1}))) \ldots \ldots \ldots (c_1 \otimes_p (c_2 \otimes_p f_j(x_1)))\right)_p$

$= (c_1 \wedge (c_2 \otimes_p IVT^{p,1}_j))(x)$.

Thus, $(T^{p,1}, \oplus, \wedge)$ forms a Vector space over $\mathbb{F}_p$.

***Remark-1:*** $T^{p,1}$ is a finite vector space (with $p^p$ functions) with dim $(T^{p,1})$ = p over the finite field $\mathbb{F}_p$. When p is composite number, $\mathbb{F}_p$ forms a commutative ring under addition and multiplication modulo p. In this case, $(T^{p,1},\oplus,\wedge)$ forms a module over $\mathbb{F}_p$ which is a commutative ring with unity. Moreover, it is a *free module* since it has a basis which is asserted in the next corollary.

**Corollary-1** The basis of $(T^{p,1},\oplus,\wedge)$ is { $IVT^{p,1}{}_j$ such that $j=p^i$ where i=0, 1, 2… p-1}.

*Proof:*

We first show linear independence of { $IVT^{p,1}{}_j$ such that $j=p^i$ where i=0, 1, 2… p-1}.

Let $a_0, a_1, a_2, \ldots, a_{p-1} \in \mathbb{F}_p$.

$a_0 \wedge IVT^{p,1}{}_{p^0} \oplus a_1 \wedge IVT^{p,1}{}_{p^1} \oplus a_2 \wedge IVT^{p,1}{}_{p^2} \oplus \ldots a_{p-1} \wedge IVT^{p,1}{}_{p^{p-1}} = 0 = IVT^{p,1}{}_0$ implies $a_i = 0 \ \forall \ i = 0,1,2,\ldots\ldots p-1$ trivially.

We now show that the set B={ $IVT^{p,1}{}_j$ such that $j=p^i$ where i=0,1,2,….p-1 } spans $T^{p,1}$ that is L(B)= $T^{p,1}$ where L(B) denotes the linear span of B.

Obviously, any linear combination of any number of functions in B will again be contained in $T^{p,1}$ i.e L(B) is contained in $T^{p,1}$.

For any $IVT^{p,1}{}_j \in T^{p,1}$, we can write it in the following form,

$IVT^{p,1}{}_j = a_0 IVT^{p,1}{}_{p^0} + a_1 IVT^{p,1}{}_{p^1} + a_2 IVT^{p,1}{}_{p^2} \ldots\ldots\ldots + a_{p-1} IVT^{p,1}{}_{p^{p-1}}$

$\in$ L(B) where $a_i \in \mathbb{F}_p$ and $j = \sum_{i=0}^{p-1} a_i p^i$. Thus, B is a linearly independent set which spans $T^{p,1}$ and is hence a basis of $T^{p,1}$.

Hence, if we know the basis elements of $T^{p,1}{}_\#$ then we can generate all the other elements of $T^{p,1}{}_\#$. Likewise the next theorem strives to establish a relationship between $T^{p,1}{}_\#$ and $T^{p+1,1}{}_\#$ i.e. between p-adic and (p+1) – adic functions.

**Theorem: 3** Let us define a function $T : B_p \cup \{ IVT^{p,1}{}_0 \} \to B_{p+1}$ as

$$T(IVT^{p,1}{}_j) = \begin{cases} IVT^{(p+1),1}{}_{(p+1)^p} & \text{if} \quad j = 0 \\ IVT^{(p+1),1}{}_{(p+1)^i} & \text{if } j = p^i \mid i = 0,1,\ldots p-1 \end{cases}$$

$B_p$ and $B_{p+1}$ are the bases of $T^{p,1}$ and $T^{p+1,1}$ respectively and $f^{(p+1),1}{}_j : \{0,1,2,\ldots\ldots p-1, p\} \to \{0,1,2,\ldots\ldots p-1, p\}$ is defined as

$f^{(p+1),1}{}_j (p) = 1$ if $f^{(p+1),1}{}_j (i) = 0 \ \forall \ i = 0,1,2,\ldots p-1$

$= 0$ if $f^{(p+1),1}{}_j (i) = 1$ for some $i = 0,1,2,\ldots p-1$.

*Then T is an Isomorphism.*

*Proof:* We first want to show that T is a linear transformation which is equivalent to showing the following

$T(a \wedge IVT^{p,1}{}_{j_1} \oplus b \wedge IVT^{p,1}{}_{j_2}) = a \wedge T(IVT^{p,1}{}_{j_1}) \oplus b \wedge T(IVT^{p,1}{}_{j_2})$, $a,b \in \mathbb{F}_p$ and $j_1 = p^{i_1}$, $j_2 = p^{i_2}$, $i_1, i_2 \in \mathbb{F}_p$

Consider the right hand side of the above equation and $x \in N$, $x = (x_n \ x_{n-1} \ldots\ldots x_1)_{p+1}$

$[a \wedge T(IVT^{p,1}_{j_1}) \oplus b \wedge T(IVT^{p,1}_{j_2})](x)$

$= [a\wedge(IVT^{(p+1),1}_{(p+1)^{i_1}}) \oplus b\wedge(IVT^{(p+1),1}_{(p+1)^{i_2}})](x)$

$= a\wedge(f_{(p+1)^{i_1}}(x_n) \ldots \ldots \ldots f_{(p+1)^{i_1}}(x_1))_{p+1} \oplus b\wedge(f_{(p+1)^{i_2}}(x_n) \ldots \ldots \ldots f_{(p+1)^{i_2}}(x_1))_{p+1}$

$= (a \otimes_{p+1} f_{(p+1)^{i_1}} \oplus_{p+1} b \otimes_{p+1} f_{(p+1)^{i_2}})(x_n) \ldots \ldots a \otimes_{p+1} f_{(p+1)^{i_1}} \oplus_{p+1} b \otimes_{p+1} f_{(p+1)^{i_2}})(x_1))_{p+1}$

$= [(a\wedge(IVT^{p+1,1}_{(p+1)^{i_1}}) \oplus b\wedge(IVT^{p+1,1}_{(p+1)^{i_2}})](x)$

$= [T(a\wedge IVT^{p,1}_{j_1} \oplus b\wedge IVT^{p,1}_{j_2})](x)$

Thus T is a linear transformation.

It is obvious that T is both injective and surjective from the definition of T. Consequently T is an Isomorphism.

This transformation T can be extended to the whole space $T^{p,1}$ to $T^{p+1,1}$ but in this case T will not be surjective and thus $T^{p,1}$ and $T^{p+1,1}$ are not isomorphic which can be seen very easily as $T^{p,1}$ has $p^p$ functions and $T^{p+1,1}_{\#}$ has $(p+1)^{p+1}$ functions but the basic structures are same for both.

The following two theorems give us an idea of the linear and bijective functions in $T^{p,1}_{\#}$

**Theorem: 4** *There are **p** number of linear functions in $T^{p,1}$. So there are $(p^p - p)$ number of non-linear functions.*

***Proof:*** For any linear function $f_{\#}$, we must have $f_{\#}(0) = 0$ and $f_{\#}(1) = i$ for $i \in \{0,1,2,\ldots\ldots p-1\}$ for a p-adic system. Consequently we would have $f_{\#}(r) = r \otimes_p f_{\#}(1)$ where $0 \leq \# \leq (p-1)$.

Thus we can have p such functions satisfying the above conditions.

**Theorem: 5** *There are p! number of bijective functions in $T^{p,1}$.*

***Proof:*** It's very clear that all the functions in $T^{p,1}$, that is all the $p^p$ functions are surjective but not all of them would be injective. For a p-adic function, there are p number of variables i.e, 0,1,2,………p-1 and an injective function say $f_j$ would have to take one of these variables again to one of these variables and hence we would have p! number of possibilities of arrangements of these variables. Thus, there would be p! number of injective functions and all of these are also surjective. Hence, there are p! number of bijective functions.

Now that we have a vector space structure, we would like to investigate whether it can be endowed with a norm which has been done below.

**Theorem: 6** $(T^{p,1}, ||.||)$ *form a normed space and the metric induced by this norm gives rise to a metric space.*

***Proof:***

We know that $T^{p,1}$ is a vector space (where p is prime) with the usual scalar multiplication and therefore it is possible to define a norm on $T^{p,1}$.

Let us define a function as,

$||.|| : T^{p,1}_{\#} \to R^+$

$||IVT^{p,1}_j|| = \begin{cases} \inf\{|IVT^{p,1}_j(x)| \text{ such that infimum is taken over x and } IVT^{p,1}_j(x) \neq 0 \text{ for } j \neq 0\} \\ 0 \quad \text{if} \quad j = 0 \end{cases}$

||.|| defines a norm and every normed space is a metric space and the metric induced by the norm is given by

$d: T^{p,1}{}_\# \times T^{p,1}{}_\# \to \mathbb{R}^+$

$d(\text{IVT}^{p,1}{}_{j_1}, \text{IVT}^{p,1}{}_{j_2}) = \|\text{IVT}^{p,1}{}_{j_1} - \text{IVT}^{p,1}{}_{j_2}\|$

$$= \begin{cases} \inf\{|(\text{IVT}^{p,1}{}_{j_1} - \text{IVT}^{p,1}{}_{j_2})(x)| \text{ such that infimum is taken over } x \\ \qquad \text{and } |(\text{IVT}^{p,1}{}_{j_1} - \text{IVT}^{p,1}{}_{j_2})(x)| \neq 0 \text{ for } j_1 \neq j_2\} \\ 0 \qquad\qquad\qquad\qquad\quad \text{if} \qquad\qquad\qquad j_1 = j_2 \end{cases}$$

Then Clearly $d(\text{IVT}^{p,1}{}_{j_1}, \text{IVT}^{p,1}{}_{j_2}) \geq 0$ and $d(\text{IVT}^{p,1}{}_{j_1}, \text{IVT}^{p,1}{}_{j_2}) = 0$ iff $j_1 = j_2$ $\forall$ $\text{IVT}^{p,1}{}_{j_1}, \text{IVT}^{p,1}{}_{j_2} \in T^{p,1}{}_\#$.

$d(\text{IVT}^{p,1}{}_{j_1}, \text{IVT}^{p,1}{}_{j_2}) = d(\text{IVT}^{p,1}{}_{j_2}, \text{IVT}^{p,1}{}_{j_1})$ $\forall$ $\text{IVT}^{p,1}{}_{j_1}, \text{IVT}^{p,1}{}_{j_2} \in T^{p,1}{}_\#$.

For $\text{IVT}^{p,1}{}_{j_1}, \text{IVT}^{p,1}{}_{j_2}, \text{IVT}^{p,1}{}_{j_3} \in T^{p,1}$,

$d(\text{IVT}^{p,1}{}_{j_1}, \text{IVT}^{p,1}{}_{j_3}) \leq d(\text{IVT}^{p,1}{}_{j_1}, \text{IVT}^{p,1}{}_{j_2}) + d(\text{IVT}^{p,1}{}_{j_2}, \text{IVT}^{p,1}{}_{j_3})$

which follows from the inequality,

$\inf\{|(IVT^{p,1}{}_{j_1} - IVT^{p,1}{}_{j_3})(x)|\} \leq \inf\{|(IVT^{p,1}{}_{j_1} - IVT^{p,1}{}_{j_2})(x)|\} + \inf\{|(IVT^{p,1}{}_{j_2} - IVT^{p,1}{}_{j_3})(x)|\}$.

Thus, $(T^{p,1}, d)$ forms a metric space.

Note: In the definition of the metric, if the condition " $|(IVT^{p,1}{}_{j_1} - IVT^{p,1}{}_{j_2})(x)| \neq 0$ for $j_1 \neq j_2$" is dropped then it defines a pseudo metric.

So far we have achieved some ring and vector space structure on $T^{p,1}$. Let us make an effort to achieve differential operator on $T^{p,1}$.

## 4. Notion of differentiability on $(T^{p,1}, \oplus, \wedge)$

To define an operator, D as a Differential operator, the following fundamental rules should be satisfied.

1) *Linearity:* $D(IVT^{p,1}{}_{j_1} + IVT^{p,1}{}_{j_2}) = D(IVT^{p,1}{}_{j_1}) + D(IVT^{p,1}{}_{j_2})$

2) *Homogeneity*: $D(k \wedge IVT^{p,1}{}_{j_1}) = k\, D(IVT^{p,1}{}_{j_1})$

3) *Leibnitz Rule*: $D(IVT^{p,1}{}_{j_1} \cdot IVT^{p,1}{}_{j_2}) = IVT^{p,1}{}_{j_1} \cdot D(IVT^{p,1}{}_{j_2}) + IVT^{p,1}{}_{j_2} \cdot D(IVT^{p,1}{}_{j_1})$

Now let us make an attempt to have an operator D on $(T^{p,1}, \oplus, \wedge)$ keeping in mind the usual notion of differentiation on continuous domain.

In the case of functions defined on a continuous domain, derivability of a function is meant for the rate at which a function is increasing/decreasing. To define the notion of differentiation on discrete domain $\mathbb{N}_0$, let us first introduce the concept of neighbourhood. The neighbourhood of a point in a continuous domain is defined as $N(x_0, r) = \{x_0 \in X: d(x, x_0) < r\}$ where d denotes the metric defined on the set X.

Now, for $x_0 \in \mathbb{N}_0$, the neighbourhood of $x_0$ consisting of discrete points boils down to the following

$N(x_0, r) = \{x \in \mathbb{N}_0: |x - x_0| < r\} = \{x_0 - r + 1, \ldots\ldots\ldots x_0 + r - 1\}; r > 1, r \in \mathbb{N}_0$

So we are in position to define the operator D on $(T^{p,1}{}_\#, \oplus, \wedge)$. We first define a left derivative for an $IVT^{p,1}{}_j$ at an arbitrary point $c \in \mathbb{N}_0$ as the following:

Let $LD(IVT^{p,1}_j)(c)$ denote the left derivative of the function $IVT^{p,1}_j$ at the point c and

$$LD(IVT^{p,1}_j)(c) = \min_{x: x \in N(c,\varepsilon)} \left\{ \frac{IVT^{p,1}_j(x) - IVT^{p,1}_j(c)}{(x-c)} \right\}$$

and

$RD(IVT^{p,1}_j)(c)$ denote the right derivative of the function $IVT^{p,1}_j$ at the point c and

$$RD(IVT^{p,1}_j)(c) = \max_{x: x \in N(c,\varepsilon)} \left\{ \frac{IVT^{p,1}_j(x) - IVT^{p,1}_j(c)}{(x-c)} \right\}.$$

If both the right and left derivative of the function $IVT^{p,1}_j$ exist at the point c and are equal, then we say that $IVT^{p,1}_j$ is differentiable at the point c and the derivative at c is equal to $D(IVT^{p,1}_j)(c)$.

To accept D, as a differential operator we need to check the following:

*Linearity:*

Let $IVT^{p,1}_{j_1}, IVT^{p,1}_{j_2} \in T^{p,1}$ and suppose are derivable at c.

$$LD(IVT^{p,1}_{j_1} + IVT^{p,1}_{j_2})(c) = \min_{x: x \in N(c,\varepsilon)} \left\{ \frac{(IVT^{p,1}_{j_1} + IVT^{p,1}_{j_2})(x) - (IVT^{p,1}_{j_1} + IVT^{p,1}_{j_2})(c)}{(x-c)} \right\}$$

$$= \min_{x: x \in N(c,\varepsilon)} \left\{ \frac{(IVT^{p,1}_{j_1}(x) - IVT^{p,1}_{j_1}(c)) + (IVT^{p,1}_{j_2}(x) - IVT^{p,1}_{j_2}(c))}{(x-c)} \right\}$$

$$= \min_{x: x \in N(c,\varepsilon)} \left\{ \frac{IVT^{p,1}_{j_1}(x) - IVT^{p,1}_{j_1}(c)}{(x-c)} \right\} + \min_{x: x \in N(c,\varepsilon)} \left\{ \frac{IVT^{p,1}_{j_2}(x) - IVT^{p,1}_{j_2}(c)}{(x-c)} \right\}$$

$$= LD(IVT^{p,1}_{j_1})(c) + LD(IVT^{p,1}_{j_2})(c)$$

*Leibnitz Rule:*

$$LD(IVT^{p,1}_{j_1} \cdot IVT^{p,1}_{j_2})(c) = \min_{x: x \in N(c,\varepsilon)} \left\{ \frac{(IVT^{p,1}_{j_1} IVT^{p,1}_{j_2})(x) - (IVT^{p,1}_{j_1} IVT^{p,1}_{j_2})(c)}{(x-c)} \right\}$$

$$= \min_{x: x \in N(c,\varepsilon)} \left\{ \frac{IVT^{p,1}_{j_1}(x)IVT^{p,1}_{j_2}(x) - IVT^{p,1}_{j_1}(x)IVT^{p,1}_{j_2}(c) + IVT^{p,1}_{j_1}(x)IVT^{p,1}_{j_2}(c) - IVT^{p,1}_{j_1}(c)IVT^{p,1}_{j_2}(c)}{(x-c)} \right\}$$

$$= \min_{x: x \in N(c,\varepsilon)} \left\{ \frac{IVT^{p,1}_{j_1}(x)\left[IVT^{p,1}_{j_2}(x) - IVT^{p,1}_{j_2}(c)\right] + \left[IVT^{p,1}_{j_1}(x) - IVT^{p,1}_{j_1}(c)\right]IVT^{p,1}_{j_2}(c)}{(x-c)} \right\}$$

$$= \min_{x: x \in N(c,\varepsilon)} \{IVT^{p,1}_{j_1}(x)\} \, LD(IVT^{p,1}_{j_2})(c) + IVT^{p,1}_{j_2}(c) \cdot LD(IVT^{p,1}_{j_1})(c)$$

Homogeneity does not hold from the fact that

$$LD(k^\wedge IVT^{p,1}_j)(c) = \min_{x: x \in N(c,\varepsilon)} \left\{ \frac{k^\wedge IVT^{p,1}_j(x) - k^\wedge IVT^{p,1}_j(c)}{(x-c)} \right\}$$

$$= \min_{x: x \in N(c,\varepsilon)} \left\{ \frac{k^\wedge IVT^{p,1}_j(x) - k^\wedge IVT^{p,1}_j(c)}{(x-c)} \right\}$$

$= LD(IVT^{p,1}_{j_1})(c)$ for some $j_1$.

Let us illustrate with an example as follows:

Consider, $IVT^{2,1}_2(x) = x$

Let c be any arbitrary natural number,

$$LD(IVT^{2,1}{}_2)(c) = \min_{x:x\in N(c,\varepsilon)} \left\{ \frac{IVT^{2,1}{}_2(x) - IVT^{2,1}{}_2(c)}{(x-c)} \right\}$$

$$= \min_{x:x\in N(c,\varepsilon)} \left\{ \frac{x-c}{(x-c)} \right\} = 1$$

$$RD(IVT^{2,1}{}_2)(c) = \max_{x:x\in N(c,\varepsilon)} \left\{ \frac{IVT^{2,1}{}_2(x) - IVT^{2,1}{}_2(c)}{(x-c)} \right\}$$

$$= \max_{x:x\in N(c,\varepsilon)} \left\{ \frac{x-c}{(x-c)} \right\} = 1$$

Therefore $IVT^{2,1}{}_2$ is everywhere differentiable as both $LD(IVT^{2,1}{}_2)(c)$ and $RD(IVT^{2,1}{}_2)(c)$ exist and are equal and the derivative is 1.

$IVT^{2,1}{}_1(x) = (2^s - 1) - x$ where x has a s-bit representation

$$LD(IVT^{2,1}{}_1)(c) = \min_{x:x\in N(c,\varepsilon)} \left\{ \frac{IVT^{2,1}{}_1(x) - IVT^{2,1}{}_1(c)}{(x-c)} \right\}$$

$$= \min_{x:x\in N(c,\varepsilon)} \left\{ \frac{(2^{s_1}-1) - x - [(2^{s_2}-1) - c]}{(x-c)} \right\} = \min_{x:x\in N(c,\varepsilon)} \left\{ \frac{(2^{s_1} - 2^{s_2})}{(x-c)} - 1 \right\} \to -1$$

$$RD(IVT^{2,1}{}_1)(c) = \max_{x:x\in N(c,\varepsilon)} \left\{ \frac{IVT^{2,1}{}_1(x) - IVT^{2,1}{}_1(c)}{(x-c)} \right\}$$

$$= \max_{x:x\in N(c,\varepsilon)} \left\{ \frac{(2^{s_1}-1) - x - [(2^{s_2}-1) - c]}{(x-c)} \right\} = \max_{x:x\in N(c,\varepsilon)} \left\{ \frac{(2^{s_1} - 2^{s_2})}{(x-c)} - 1 \right\} \to +\infty$$

As $LD(IVT^{2,1}{}_1)(c)$ and $RD(IVT^{2,1}{}_1)(c)$ are not equal therefore $IVT^{2,1}{}_1$ is nowhere differentiable.

Now we have the basic foundation which sets the stage for further research. We next make an attempt to understand the dynamism of these functions.

## 5. *Dynamical system of IVT*

We introduce a dynamical system of the IVT to gauge the behaviour of these functions over time.

$T : \mathbb{N}_0 \times \mathbb{N}_0 \to \mathbb{N}_0$

$T(n, x) = T_n(x)$

where $T_n(x) = (IVT^{p,1}{}_j)^n(x) = \underbrace{[(IVT^{p,1}{}_j)(IVT^{p,1}{}_j)\ldots\ldots(IVT^{p,1}{}_j)]}_{n\ times}(x)$

$(\mathbb{N}_0, x)$ is a semi-group acting on the space $\mathbb{N}_0$ and $(\mathbb{N}_0, T)$ is a discrete dynamical system.

Since we have, $T_{n_1 \times n_2} = (IVT^{p,1}{}_j)^{n_1 \times n_2} = \underbrace{[(IVT^{p,1}{}_j)(IVT^{p,1}{}_j)\ldots\ldots(IVT^{p,1}{}_j)]}_{n_1 \times n_2\ times} =$

$\underbrace{[(IVT^{p,1}{}_j)(IVT^{p,1}{}_j)\ldots\ldots(IVT^{p,1}{}_j)]}_{n_1\ times} \times \underbrace{[(IVT^{p,1}{}_j)(IVT^{p,1}{}_j)\ldots\ldots(IVT^{p,1}{}_j)]}_{n_2\ times}$

$= T_{n_1} \times T_{n_2}$

$T(n, x)$ is the evolution function and n is the evolution parameter of the dynamical system.

Let $X_0$ be the initial state and $\mathbb{N}_0$ is the state/phase space. Corresponding to each p and j, we get a different dynamical system.

The orbit of $x_0$ is a set of points denoted by
$\gamma(x_0) = \{x_n | n \in \mathbb{N}_0, x_{n+1} = (IVT^{p,1}_j)(x_n)\}$
$= \{x_0, (IVT^{p,1}_j)(x_0), (IVT^{p,1}_j)^2(x_0), (IVT^{p,1}_j)^3(x_0), \ldots \ldots \ldots\}$

The orbits corresponding to a particular IVT can aid us in gaining more insight on the behaviour of the IVTs.

For instance, the orbit of $IVT^{2,1}_1$ around $x_0$ is given by

$\gamma(x_0) = \{x_0, (IVT^{2,1}_1)(x_0), (IVT^{2,1}_1)^2(x_0), (IVT^{2,1}_1)^3(x_0), \ldots \ldots \ldots\}$
$\gamma(0) = \{0,1\}, \gamma(1) = \{0,1\}, \gamma(2) = \{2,1,0\}, \gamma(3) = \{3,1,0\}, \gamma(4) = \{4,3,1,0\}, \gamma(5) = \{5,2,1,0\},$
$\gamma(6) = \{6,1,0\}, \gamma(7) = \{7,1,0\}, \gamma(15) = \{15,1,0\}$

It's worth noting that the orbits of the Merseene numbers are 3-point sets which consist of the number itself and 0 and 1 in case of $IVT^{2,1}_1$ and the rest converge to 0 through the Merseene numbers. The orbits basically give us the path/trajectory of convergence of the functions. $IVT^{2,1}_1$ is a Collatz like function as it converges to 0.

For $IVT^{2,1}_2(x) = x, \gamma(x_0) = \{x_0, (IVT^{2,1}_2)(x_0), (IVT^{2,1}_2)^2(x_0), (IVT^{2,1}_2)^3(x_0), \ldots \ldots \ldots\} = \{x_0\}$

The orbit of any steady state equilibrium/fixed point will be the point itself and likewise we can characterize other functions.

Owing to the abundance of IVTs in a p-adic system, the need for classification of these functions arises and one way of doing it could be through orbit classification.

### *6. Conclusion and Future research discussion:*

In summary, we have made an effort to have some basic mathematics like commutative ring and vector space structures in $T^{p,1}$. But as we have seen so far we have not been able to achieve field structure on $T^{p,1}$. Our effort is to achieve field structure or at least division algebra structure. In the near future we will explore the dynamics of these IVTs in terms of dynamical system and through topological dynamics in the set of natural numbers.

*Acknowledgement:* Authors are grateful to **Dr. Sudhakar Sahoo** and **Dr. Prasanta Sinha** for their constructive suggestions for this research work.*References:*

[1] Hassan, Sk. S. et al. 2010. Collatz Function like Integral Value Transformations, Alexandria Journal of Mathematics, Vol 1(2), 31-35.

[2] Wolfram, S., 1983. Statistical mechanics of Cellular Automata, Rev Mod Phys. 55,601-644.

[3] Choudhury, P. Pal et al, 2010. Theory of Carry Value Transformation (CVT) and its Application in Fractal formation, Global Journal of Computer Science and Technology, 10(14), 89-99.

[4] Choudhury, P. Pal et al, 2011. Act of CVT and EVT in the formation of number theoretic fractals, International Journal of Computational Cognition, 9(1), March 2011.